\documentclass[11pt]{article}
\usepackage{doc}
\usepackage{makeidx}

\usepackage{epsfig}
\usepackage{subfigure}
\usepackage{graphicx}
\usepackage{dcolumn}
\usepackage{bm}{

\title{On Stability of Targets for Plasma Jet Induced Magnetoinertial Fusion}

\author{ Roman Samulyak$^{1,2}$, Lina Zhang$^{1}$, Hyoungekun Kim$^{1}$ \\
\it $^1$Department of Applied Mathematics and Statistics \\
\it Stony Brook University, Stony Brook NY 11794 \\
\it Computational Science Center, \\
\it Brookhaven National Laboratory, Upton, NY, USA }

\date{April 15, 2015}
\begin{document}

\maketitle

%\keywords{plasma liner, plasma target, magneto-inertial fusion, high
%energy density matter}

\begin{abstract}
 The compression and stability of plasma targets for the plasma
 jet-induced magneto-inertial fusion (PJMIF) have been investigated
 via large scale simulations using the FronTier code capable of
 explicit tracking of material interfaces.  In the PJMIF concept, a
 plasma liner, formed by the merger of a large number of radial,
 highly supersonic plasma jets, implodes on a magnetized plasma target
 and compresses it to conditions of the fusion ignition. A multi-stage
 computational approach for simulations of the liner-target
 interaction and the compression of plasma targets has been developed
 to minimize computing time. Simulations revealed important features
 of the target compression process, including instability and
 disintegration of targets. The non-uniformity of the leading edge of
 the liner, caused by plasma jets as well as oblique shock waves
 between them, leads to instabilities during the target
 compression. By using front tracking, the evolution of targets has
 been studied in 3-dimensional simulations. Optimization studies of
 target compression with different number of jets have also been
 performed.

\end{abstract}

%------------------------------------------------------------------------------
\section{Introduction}
\label{sect:introduction}

In the Plasma Jet Induced Magneto-inertial Fusion PJMIF concept
\cite{Thio99,Hsu12}, a plasma liner, formed by the merger of a large
number of radial, highly supersonic plasma jets, implodes on a
magnetized plasma target and compresses it to conditions of the fusion
ignition.  By avoiding major difficulties associated with both the
traditional laser driven inertial confinement fusion and solid liner
driven MTF, the plasma-liner driven magneto-inertial fusion
potentially provides a low-cost and fast R\&D path towards the
demonstration of practical fusion energy.

A simplified PJMIF model was theoretically studied in \cite{Parks08},
proposing scaling laws and the analysis of target compression rates,
deconfinement times, and fusion energy gains.  A number of numerical
simulations have also been performed for the liner and target systems
in spherically-symmetric geometry. These include 1D Lagrangian
simulations \cite{Awe11}, front tracking simulations of the liner -
target interface \cite{Sam10,Kim12}, 3D SPH simulations of the
converging liner \cite{Cassibry}, and 3D simulations with resolution
of oblique shock waves and atomic processes (ionization) of the merger
of high Mach number argon jets and the formation and implosion of
liners \cite{Kim13}. The simulation effort was recently complemented
by experiments conducted at Los Alamos National Laboratory
\cite{PLX1,PLX2,PLX3}, in which a single supersonic argon plasma jet
produced by a pulsed-power-driven plasma railgun, and a merger of two
such jets were studied.
 
All previous three-dimensional simulations studies
\cite{Kim12,Cassibry} have focused only on the structure and state of
plasma liners during the merger and implosion process.  In this paper,
we report results of simulation study of a plasma target compressed by
a liner formed by the merger of 90 argon plasma jets. Simulations use
the front tracking capability of the FronTier code \cite{SamDu07},
critical for accurate resolution of large density discontinuity of the
plasma - liner interface.  Front tracking is also important for
correct description of the change of material properties across the
interface as the argon liner is described by a weakly ionized plasma
equation of state with the resolution of atomic processes
\cite{Kim12}, while the target is in a fully ionized plasma state
described by the ideal gas equation of state.  MHD processes in the
target were not included. MHD forces in the real magnetized target may
be strong, providing significant stabilizing effect. The reason to
ignore MHD effects was partly motivated by current plans to perform
experiments on the compression of gas targets first, before using
magnetized targets.  While the FronTier code \cite{SamDu07} is capable
of the simulation of MHD in geometrically complex domains within the
method of front tracking, the currently implemented MHD regime, so
called the low magnetic Reynolds number approximation, is suitable for
weakly ionized plasmas but not for fully ionized plasma in the
target. The implementation of ideal MHD with front tracking will be
performed in the future.

To obtain simulation of such a multiscale process as the propagation
of free plasma jets, their merger, formation and implosion of liners,
and compression of targets, a multi-stage simulation method was
designed. It is described in detail in the next section. Simulations
were performed on a parallel supercomputer. The FronTier code
demonstrates good parallel scalability. It has been used for large
scale simulations on various platforms including USA Leadership
Computing Facilities.  FronTier was the basis of INCITE 2011 and 2012
supercomputing awards to study uncertainty quantification for
turbulent mixing and combustion.

\section{Numerical Methods}

Front tracking is a hybrid Lagrangian - Eulerian
computational method in which a lower-dimensional Lagrangian mesh,
called the interface, moves through a volume-filling Eulerian mesh and
tracks discontinuities or distinguished waves. FronTier is a
multiphysics code based on front tracking that implements various
hydrodynamic flow regimes.  FronTier is capable of robustly handling
geometrically complex interfaces and resolving their topological
changes. It supports compressible and incompressible Navier-Stokes
equations, MHD equations in the low magnetic Reynolds number
approximation \cite{SamDu07}, phase transitions \cite{RS7}, combustion
and turbulent mixing. It has been broadly used for multiple
applications including high power mercury jet targets for future
accelerators \cite{RS11,RS14,RS15}, cryogenic pellet fueling of nuclear
fusion devises \cite{RS12,RS13}, and nuclear fission applications.

The multi-stage scheme for the overall simulation is discussed in the
next section together with simulation results.

\section{Analysis of Simulation Results}

\begin{figure}
\centering
{\label{3d_target_den_05}\psfig{figure=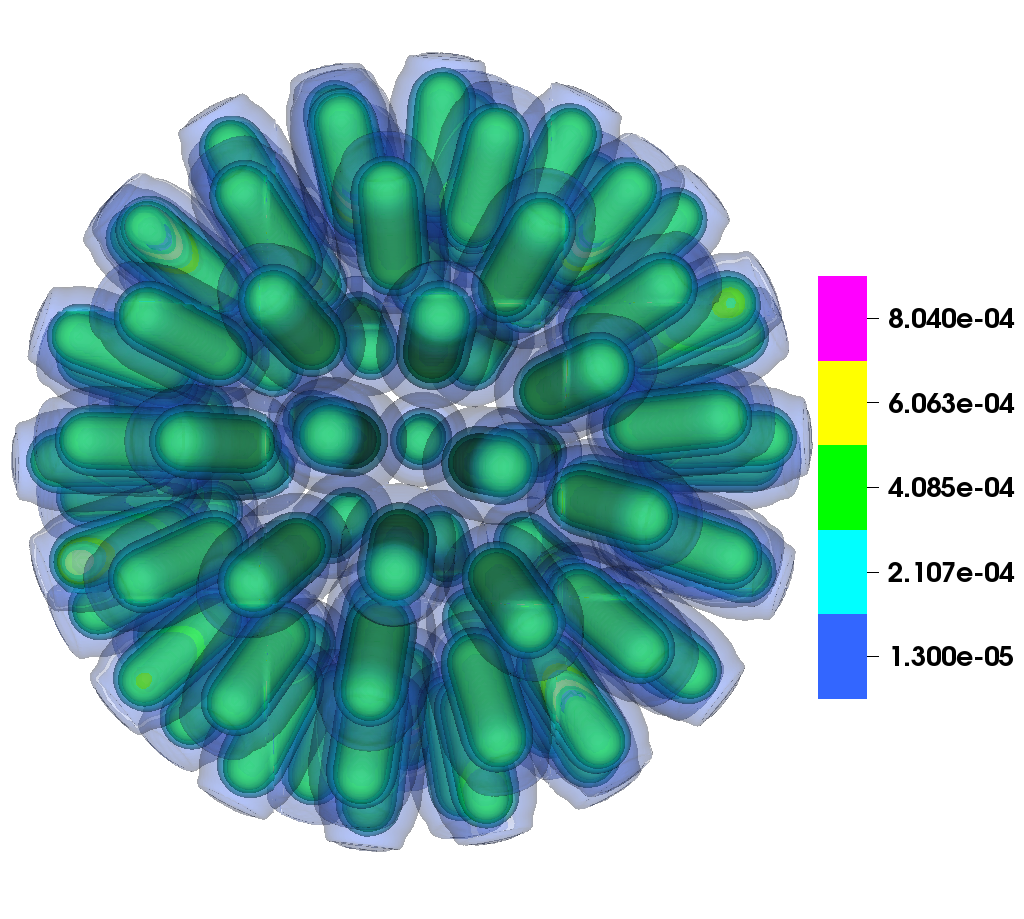,width=0.49\textwidth}}
{\label{3d_target_den_05}\psfig{figure=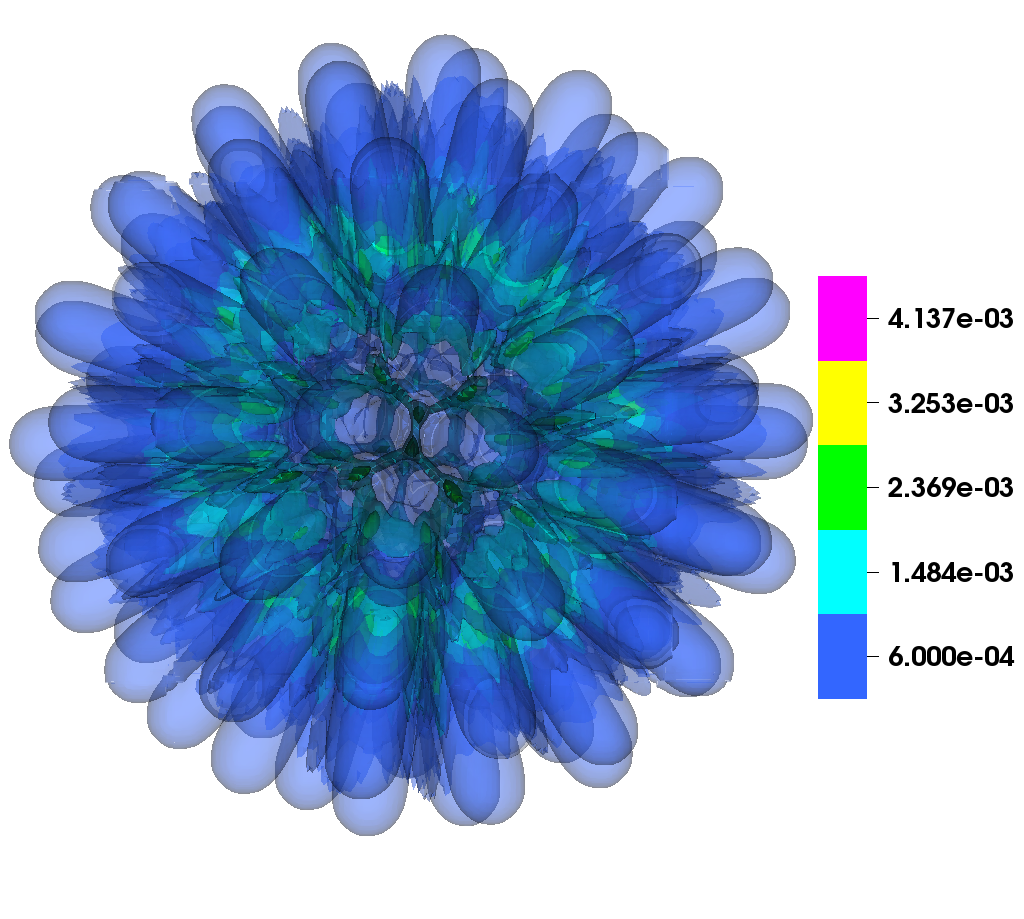,width=0.49\textwidth}}
\caption{Density contour evolution of the liner formed by 90 jets
  before interaction with the target at time 0.0097 ms and 0.0123ms.
\label{fig:3d_target_den_05}}
\end{figure}

\begin{figure}
\centering
{\label{3d_target_pre_05}\psfig{figure=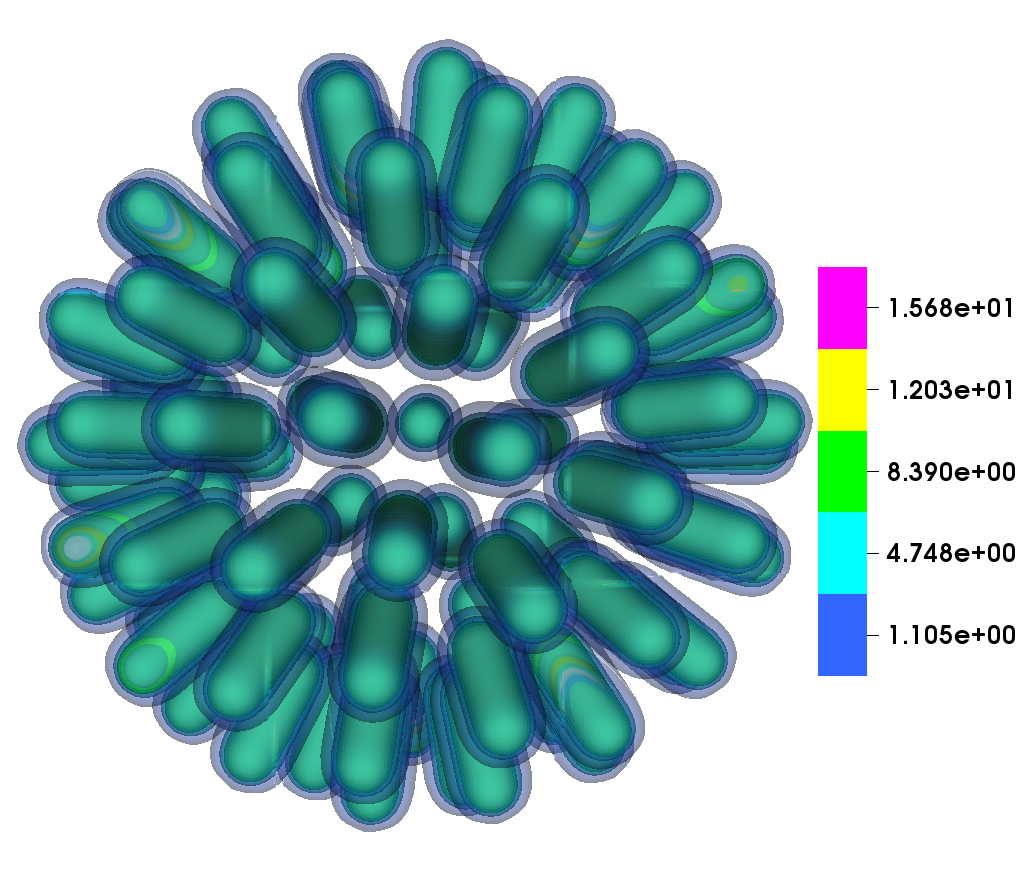,width=0.49\textwidth}}
{\label{3d_target_pre_05}\psfig{figure=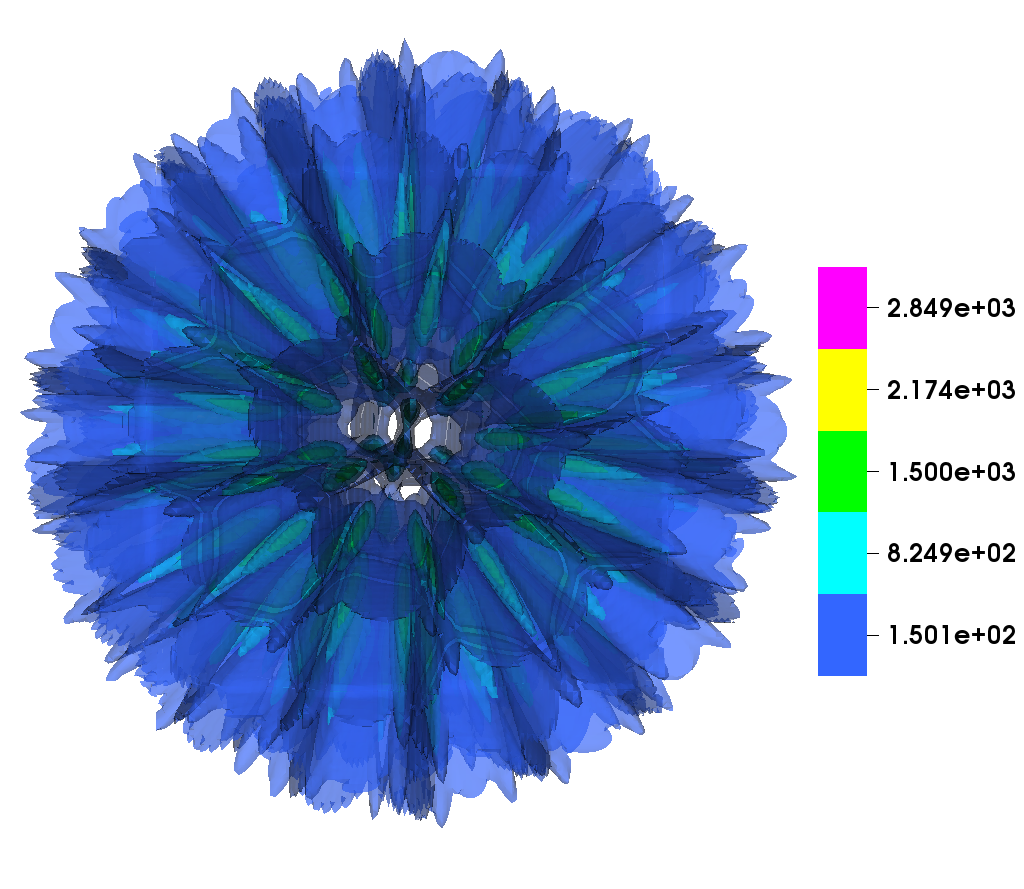,width=0.49\textwidth}}
\caption{Pressure contour evolution of the liner formed by 90 jets
  before interaction with the target at time 0.0097 ms and 0.0123ms.
\label{fig:3d_target_pre_05}}
\end{figure}

First we perform cylindrically-symmetric, 2-dimensional simulation of
the propagation of a single detached argon jet from the nozzle of the
plasma gun.  Here we use the code features that include
high-resolution hyperbolic solvers based on the Riemann problem, front
tracking, and weakly ionized plasma EOS model with atomic
processes. The argon jet has the following initial conditions: the
initial inner and outer radii are 137.2 cm and 162.7 cm, respectively,
density $\rho = 8.04\times 10^{-4}\, \mbox{g/cm}^3$, pressure $P =
18.59\, \mbox{bar}$, velocity $v = 100\, \mbox{km/s}$, and Mach number
$M = 60$.  The ambient vacuum is modeled as rarefied gas with density
$\rho_0\sim 10^{-9}~g/cm^3$ and pressure $\sim 10^{-6}$ $bar$.  The
computational mesh size is 2 mm. After obtaining the pressure,
density, and velocity profiles for the single jet before the merger,
we first find directions for 90 jets uniformly distributed in space
using Spherical Centroidal Voronoi Tessellation (SCVT), and then
initialize states of 90 jets before the merger in a 3-dimensional
code. We perform transformation from 2-dimensional cylindrical
coordinate into 3-dimensional Cartesian coordinate together with
bi-linear interpolation, which initializes the states around each jet
direction.  The initial mesh size at this stage is 5 mm which we found
to be sufficient for the resolution of the jet-merger process. This
mesh size is further refined as the simulation progresses. The target
is not included in this coarser simulation since we only need the
liner information before liner-target interaction. Figures
\ref{fig:3d_target_den_05} and \ref{fig:3d_target_pre_05} depict the
density and pressure contours before and after the jet merger. Due to
oblique shock waves, we observe redistribution of states in the
converging liner. At the later stage, the highest pressure and density
appear along the plane of interaction of the neighboring jets. This
non-uniform distribution causes the instabilities on the target. We
also observe the contours with shapes of pentagon and hexagon
determined by the location of jets.

\begin{figure}
\centering
{\label{3d_target_den_02}\psfig{figure=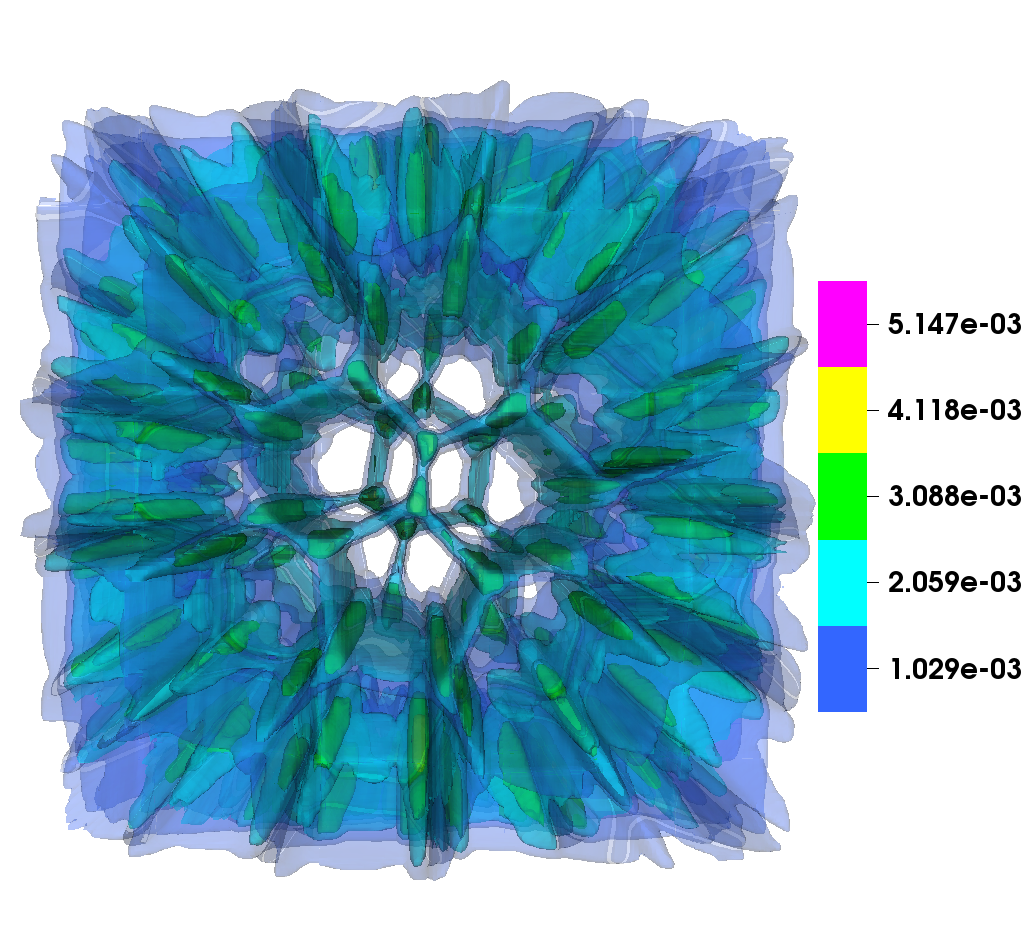,width=0.49\textwidth}}
{\label{3d_target_den_02}\psfig{figure=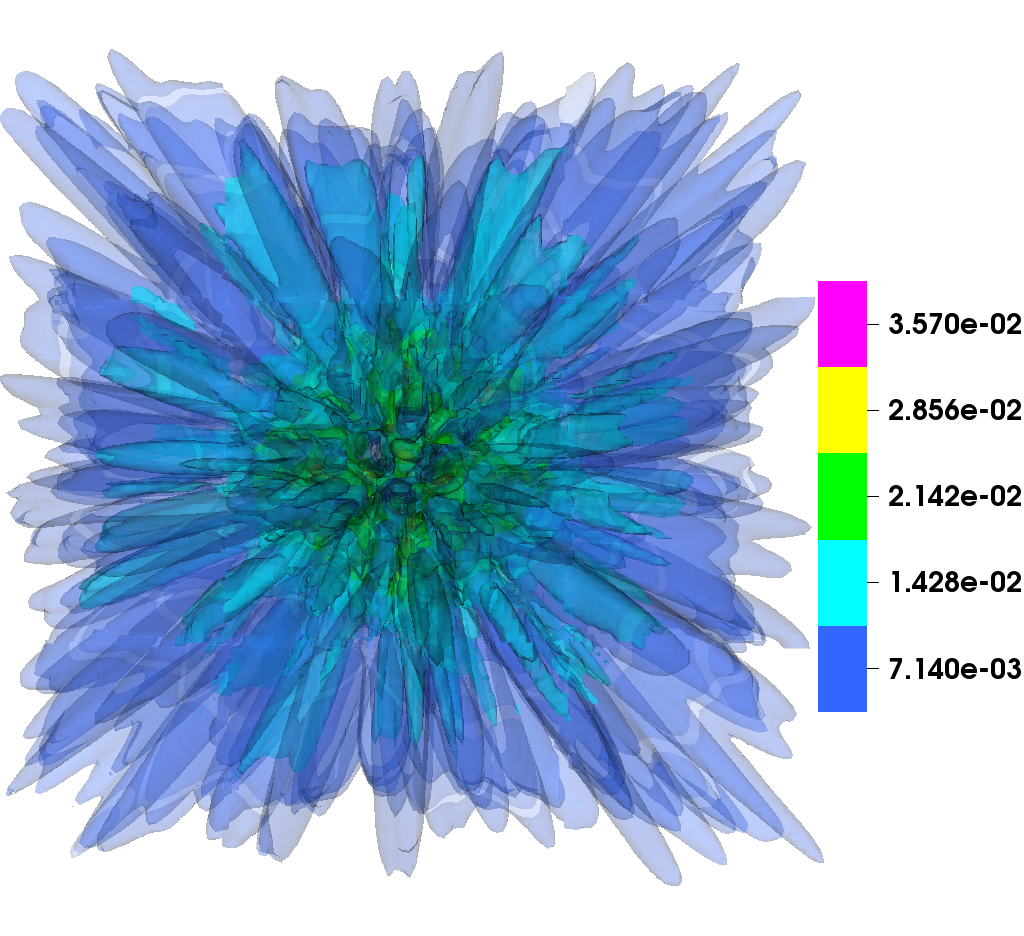,width=0.49\textwidth}}
\caption{Density contour evolution of the liner and target after their
  interaction at time 0.0124 ms and 0.0135 ms.
\label{fig:3d_target_den_02}}
\end{figure}

\begin{figure}
\centering
{\label{3d_target_pre_02}\psfig{figure=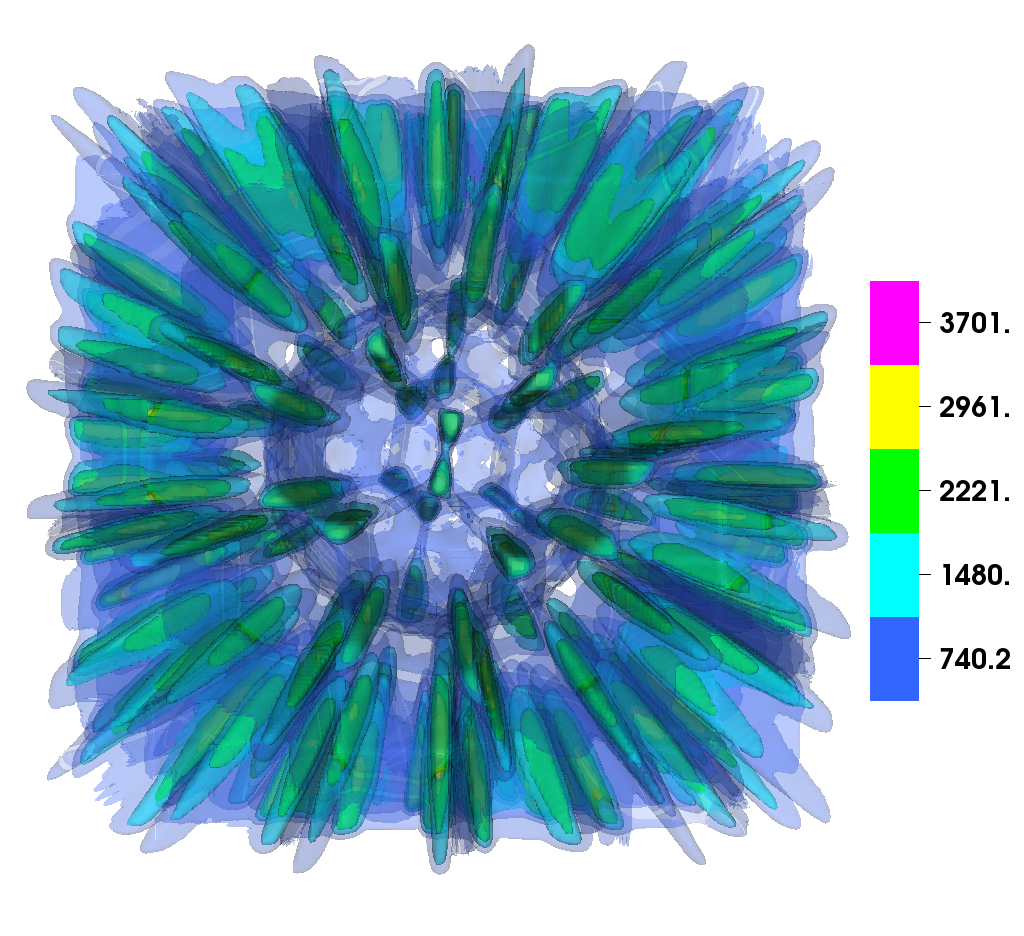,width=0.49\textwidth}}
{\label{3d_target_pre_02}\psfig{figure=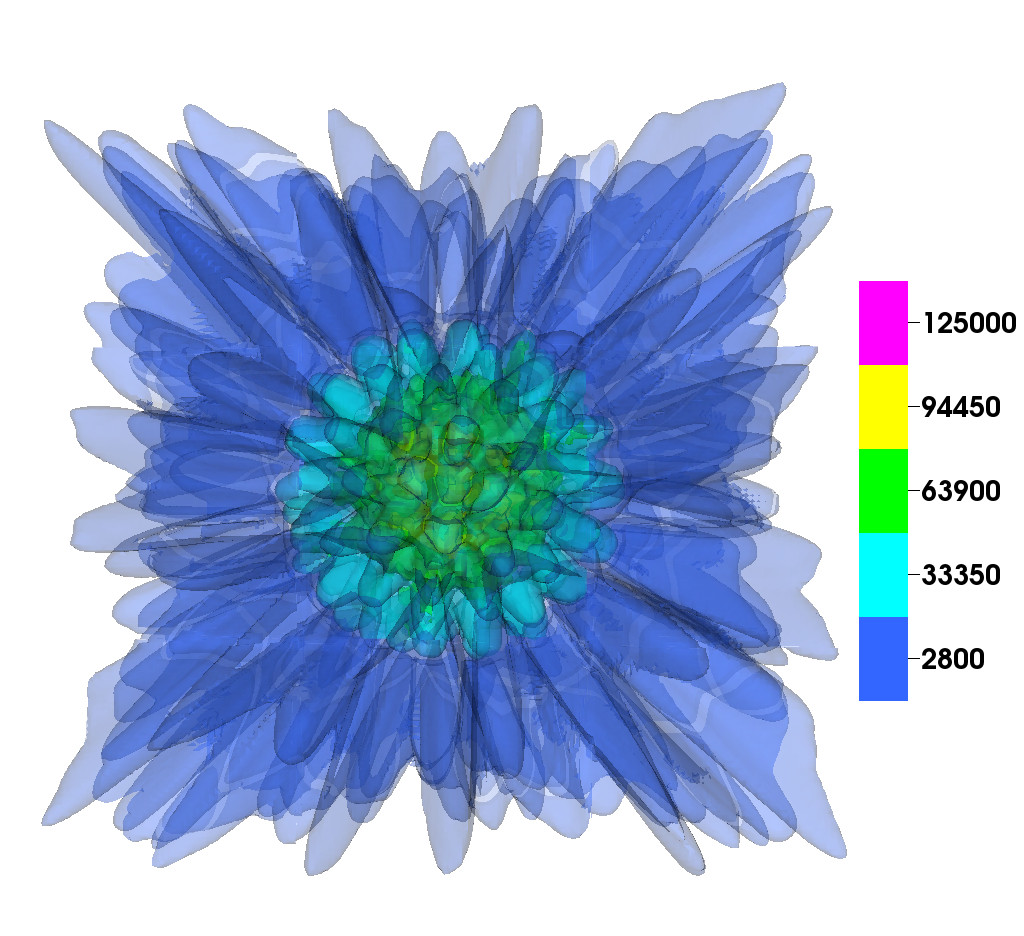,width=0.49\textwidth}}
\caption{Pressure contour evolution of the liner and the target after
  interaction at time 0.0124 ms and 0.0135 ms.
\label{fig:3d_target_pre_02}}
\end{figure}

\begin{figure}
\centering
{\label{3dpre_intfc1}\psfig{figure=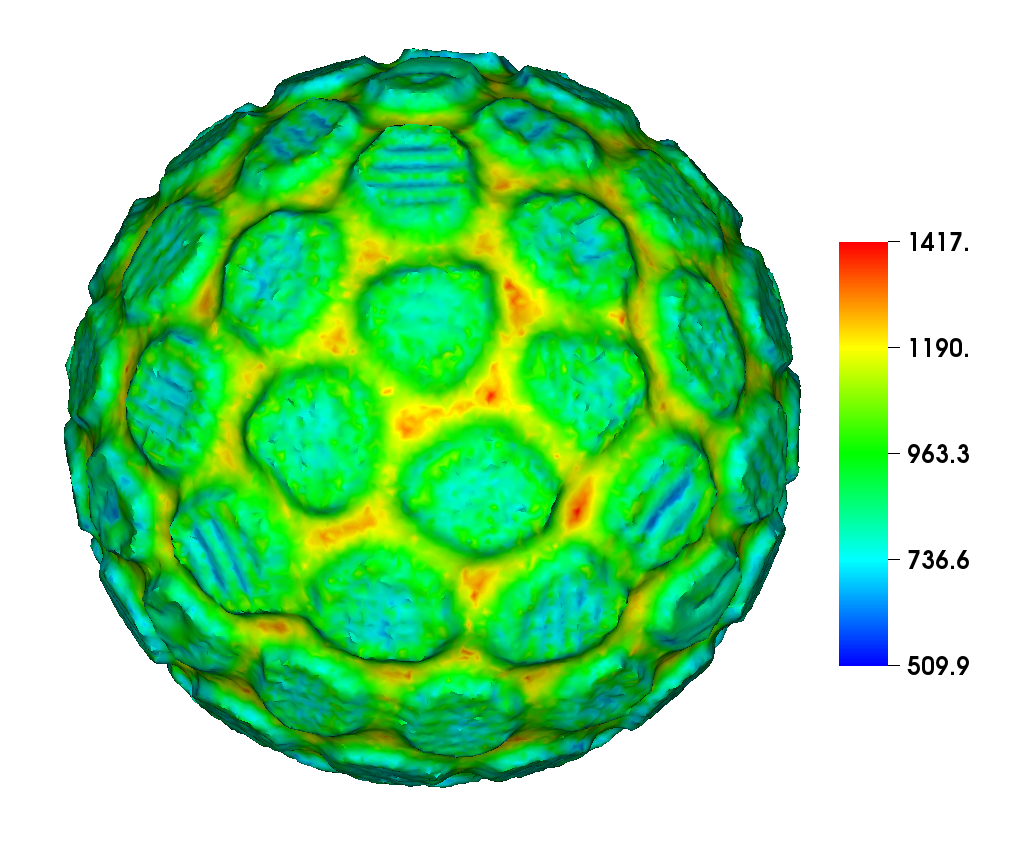,width=0.5\textwidth}}
{\label{3dpre_intfc3}\psfig{figure=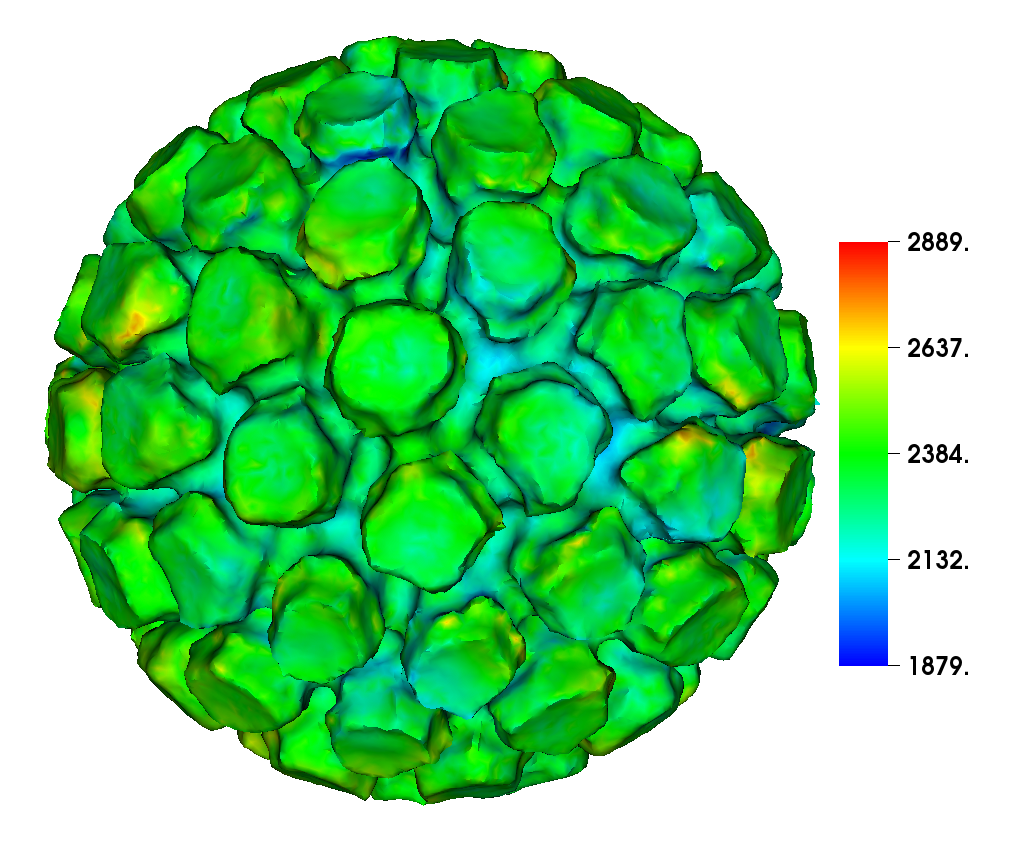,width=0.49\textwidth}}
{\label{3dpre_intfc6}\psfig{figure=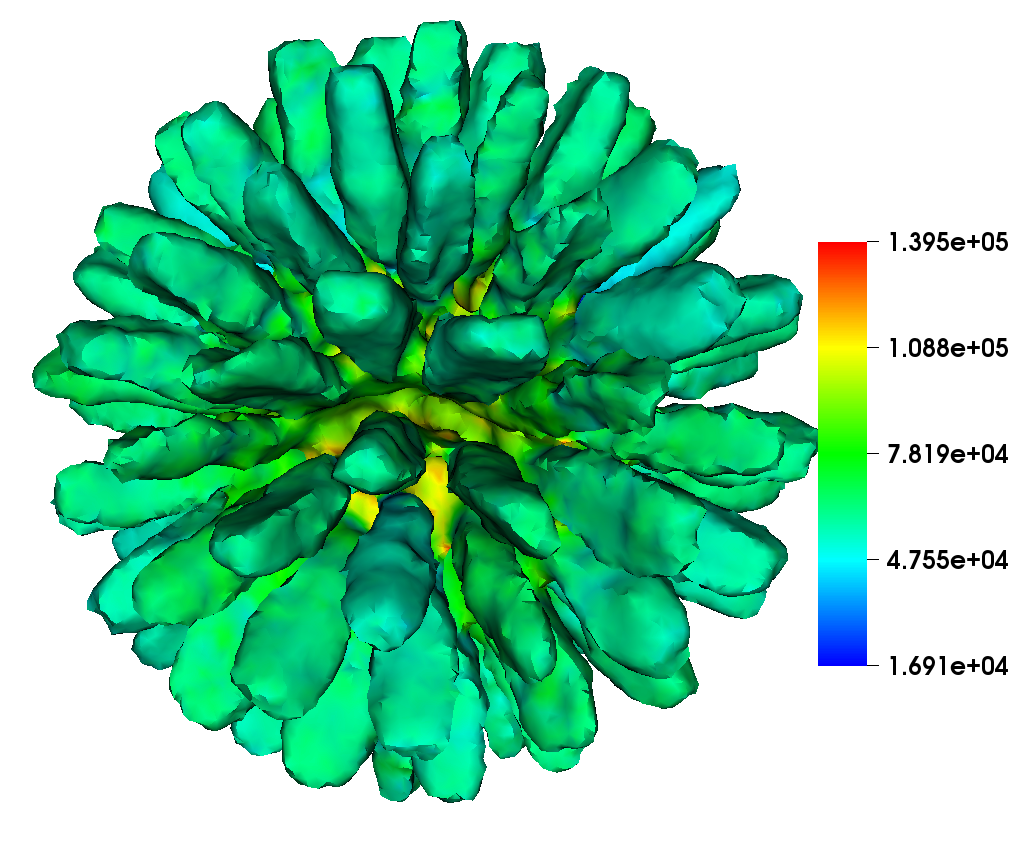,width=0.49\textwidth}}
\caption{Target evolution together with pressure distributions on the
  interface at time 0.0124 ms, 0.013 ms and 0.0135 ms.
\label{fig:3d_target_intfc}}
\end{figure}

\begin{figure}
\centering
{\label{fig:hei3d}\includegraphics[width=0.49\textwidth]{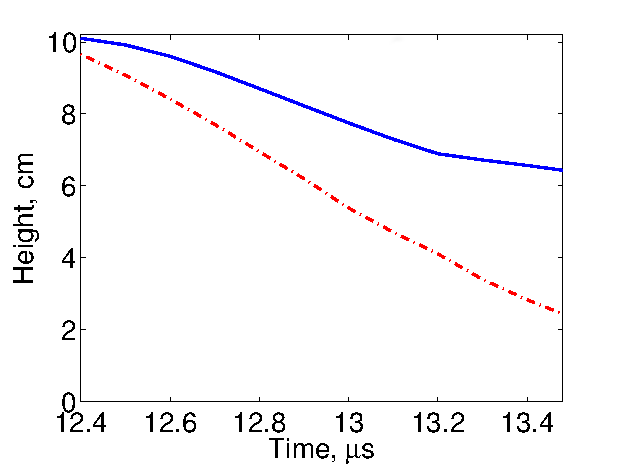}}
{\label{fig:vel3d}\includegraphics[width=0.49\textwidth]{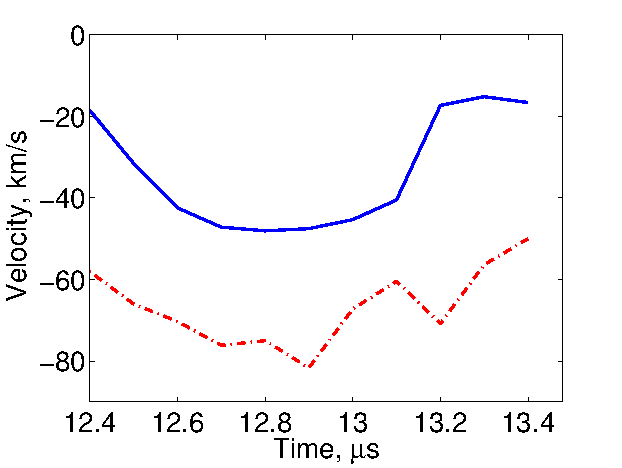}}
\caption{(a) Bubble (blue solid line) and spike (red dashed-dotted
  line) heights and (b) Bubble (blue solid line) and spike (red
  dashed-dotted line) velocities evolution from starting of
  interaction until around target breaking into fragments for
  3-dimensional simulation with mesh size as 2 mm based on 5 mm.}
\label{fig:heivel3d}
\end{figure}

Finally, we take the data of the center area from the previous
simulation when the liner still remains at some small distance off the
target, and re-initialize a refined simulation for the target
compression study.  The target initial condition is as follows:
density $\rho = 8.3\times 10^{-6}\, \mbox{g/cm}^3$, pressure $P =
640.3\, \mbox{bar}$. In order to prevent target diffusion, we set the
velocity of the target to be zero before the liner-target
interaction. As the target is compressed by a non-uniform liner, it
develops surface instabilities and even brakes in fragments at the
late stage.  The target behavior is unstable and complicated after
this stage and we currently only focus on the properties of the target
before it brakes into fragments.  Figure \ref{fig:3d_target_den_02}
and Figure \ref{fig:3d_target_pre_02} depict density and pressure
contours evolution in the center region including liner and
target. Figure \ref{fig:3d_target_pre_02} shows the interaction
between the liner and target with formation of bubbles and spikes on
the target at later stage. Here spikes are inward pointing toward the
target and bubbles are outward pointing toward the liner.  The region
with higher density and pressure along the plane of interaction of the
neighboring jets compresses the target with higher rate and bubbles
and spikes are obtained. The instabilities are amplified with time. In
order to inspect the evolution of the target more clearly, we present
the evolution of target together with pressure distribution on the
interface in Figure \ref{fig:3d_target_intfc}. The maximum pressure
appears on the spikes, in the region of interaction of the neighboring
jets which is also the region of the maximum pressure for the liner
due to oblique shock waves. The target finally breaks because of this
uneven pressure distribution.  Figure \ref{fig:heivel3d} shows the
properties of bubbles and spikes. The bubble and spike heights keep
increasing, and the terminal bubble velocity becomes quasi-constant
around $12.6~\mu s\to 13.1~\mu s$, before the breakup of the target.

\begin{figure}
\centering
\epsfig{figure=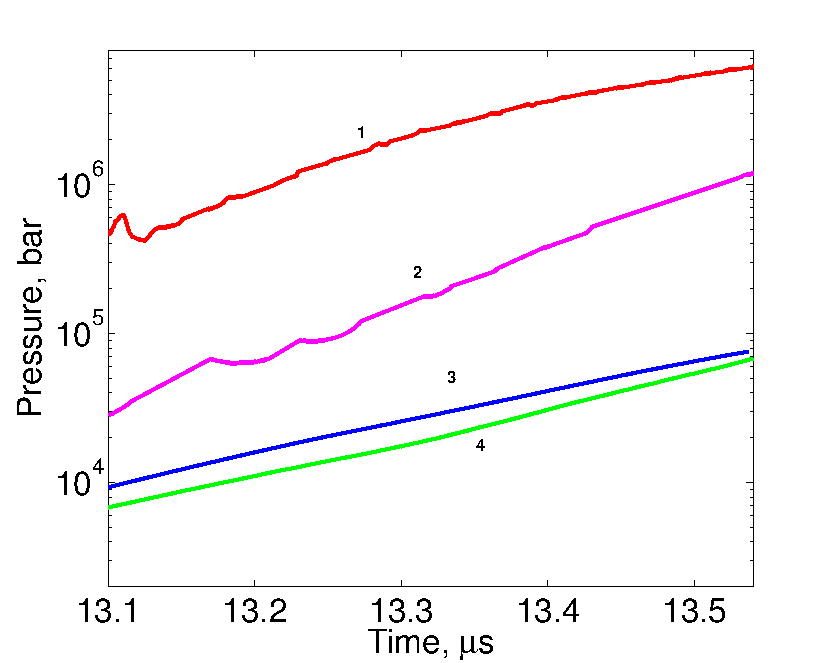,width=0.6\columnwidth}
\caption{Average pressure in target for (1) 3-dimensional uniform case
  (1-dimenstional spherical geometry); (2) 2-dimensional uniform case
  (1-dimenstional cylindrical geometry); (3) 3-dimensional case (90
  jets); (4) 2-dimensional case (16 jets).}
\label{fig:pressure_difdim}
\end{figure}

Finally, we perform 1-dimensional spherical symmetric simulation
corresponding to a 3-dimensional uniform liner to compare with the
full 3-dimensional simulation and quantify the role of
non-uniformities and instabilities of the liner and target.  The
initial conditions, obtained by the averaging of the 3-dimensional
data in angular coordinates, are as follows: the inner and outer radii
are 137.2 cm and 162.7 cm respectively, the density $\rho =
1.744\times 10^{-5}\, \mbox{g/cm}^3$, pressure $P = 0.421\,
\mbox{bar}$, velocity $v = 100\, \mbox{km/s}$, and Mach number $M =
60$.  Figure \ref{fig:pressure_difdim} shows the average pressure in
the target for different cases. Note that we only focus on the time
range of the target compression before fragmentation. The pressures of
the 3-dimensinal (90 jets) case and 2-dimensional (16 jets) case are
very close to each, around $P = 7.5e4\, \mbox{bar}$ and $P = 7.1e4\,
\mbox{bar}$ respectively at the end of this time range. The pressure
of uniform cases is always higher as expected because of the impact of
oblique shock waves for jet case. The 3-dimensinal uniform case
(1-dimenstional spherical geometry) is around $P = 6.3\times10^6\,
\mbox{bar}$ while the 2-dimensinal uniform case (1-dimenstional
cylindrical geometry) is around $P = 1.3\times10^6\, \mbox{bar}$. The
pressure of 3-dimensinal uniform case is almost 80 times higher than
that of the 90 jets case. Similar simulations of a self-implosion of
liners (without a target) produce the difference of stagnation
pressure between a 3-dimensional simulation and the corresponding
1-dimensional uniform problem of about 50 times.

We would like to emphasize that current simulations were performed at
conditions compatible with capabilities of the experimental facility
at Los Alamos National Laboratory. The achievable pressures and
temperatures in targets are well below the fusion ignition. Therefore
we do not comment on the fusion energy gain in this work.

\section{Conclusions}

In this paper, we investigated the compression and stability of plasma
targets for the plasma jet induced magneto-inertial fusion (PJMIF) via
large scale simulations using the FronTier code capable of explicit
tracking of material interfaces.  A multi-stage computational approach
for simulations of the liner-target interaction and the compression of
plasma targets has been developed to minimize computing
time. Simulations involve the propagation of a single supersonic argon
plasma jet, the merger of 90 jets and the formation of a plasma liner,
the implosion of the liner, and the compression of a plasma
target. Simulation show the formation and evolution of oblique shock
waves during the jet merger process, consistent with previous
studies. These shock waves reduce the average Mach number of the liner
and its ability to compress the target, and determine to a large
extent the nonuniform properties of the imploding liner.  Simulations
revealed important features of the target compression process,
including instability and disintegration of targets. The
non-uniformity of the leading edge of the liner, caused by plasma jets
as well as oblique shock waves between them, leads to instabilities
during the target compression.  Optimization studies of target
compression with different number of jets have also been performed.

\section{Acknowledgement} Research was partially supported by a DOE OFES grant.

\end{document}